\documentclass[twocolumn,prb,amsmath,superscriptaddress,showpacs]{revtex4}
\usepackage{graphics}
\usepackage{graphicx}
\usepackage{dcolumn} 
\usepackage{bm}
\usepackage{epsfig}

\def\byio*{Ba$_2$YIrO$_6$}
\def\syio*{Sr$_2$YIrO$_6$}
\newcommand{\bR}{\mathbf{R}}

\def\kha*{Khaliullin}
\def\s*{$|s\rangle$}
\def\t*{$|t_\alpha\rangle$}
\def\se*{$s$}
\def\te*{$t_{\alpha}$}
\def\jo*{$j_{1/2}$}
\def\jt*{$j_{3/2}$}

\begin{document}
\title{On the possibility of excitonic magnetism in Ir double perovskites}
\author{K. Pajskr}
\affiliation{Charles University in Prague, Faculty of Mathematics and Physics, Department of Condensed Matter Physics,
Ke Karlovu 5, 121 16 Prague 2, Czech Republic}

\author{P. Nov\'ak}
\affiliation{Institute of Physics,
Academy of Sciences of the Czech Republic, Cukrovarnick\'a 10,
162 53 Praha 6, Czech Republic}

\author{V. Pokorn\'y}
\affiliation{Institute of Physics,
Academy of Sciences of the Czech Republic, Na Slovance 2,
182 21 Praha 8, Czech Republic}
\affiliation{Theoretical Physics III, Center for Electronic Correlations and Magnetism, Institute of Physics, University of Augsburg,
D-86135 Augsburg, Germany}
\author{J. Koloren\v{c}}
\affiliation{Institute of Physics,
Academy of Sciences of the Czech Republic, Na Slovance 2,
182 21 Praha 8, Czech Republic}

\author{R. Arita} 
\affiliation{RIKEN Center for Emergent Matter Science (CEMS), Wako, Saitama 351-0198, Japan}
\affiliation{JST ERATO Isobe Degenerate $\pi$-Integration Project, 
Advanced Institute for Materials Research (AIMR),
Tohoku University, Sendai, Miyagi 980-8577, Japan}

\author{J. Kune\v{s}}
\email{kunes@fzu.cz}
\affiliation{Institute of Physics,
Academy of Sciences of the Czech Republic, Na Slovance 2,
182 21 Praha 8, Czech Republic}

\date{\today}

\begin{abstract}
We combine several numerical and semi-analytical methods to study the $5d$ double perovskites
\syio* and \byio* which were recently proposed to exhibit excitonic
magnetism. Starting from the density functional theory and constrained 
random phase approximation we construct effective multi-band Hubbard models.
These are analyzed by means of static and dynamical mean-field theories 
and strong coupling expansion. We find both materials to be insulators, but, 
contrary to the experimental claims, with a large spin gap of several hundreds meV
preventing formation of an ordered state at low temperature.    
\end{abstract}
\maketitle
\section{Introduction}
The discovery of new spin-orbit related phenomena and states of matter
lead to an immense increase of interest in materials containing $5d$ elements, Ir and Os in particular.~\cite{kim08}
Among the new effects, \kha*~\cite{kha13} recently proposed a possibility of excitonic condensation
in $d^4$ materials with cubic local symmetry.
The singlet atomic ground state  \s* promises no interesting low-energy physics.
However, if the energy of the lowest excitation, a magnetic moment carrying triplet \t*, is sufficiently
small, the inter-atomic exchange due to electron hopping may lead to formation of an ordered state - excitonic magnet. 
Thanks to the singlet-triplet exchange processes, the \te* excitations can be viewed
as mobile quasi-particles propagating on the singlet background with a non-trivial dispersion.
If the the bottom of  this dispersion touches the singlet level 
a phase transition (Bose-Einstein condensation) takes place. 
In some systems pairwise creation and annihilation of \te*'s on the neighboring atoms plays a role
similar to the \te* hopping.

Realization of the \kha*'s proposal~\cite{kha13} is an interesting challenge since the spin-orbit coupling must be 
strong enough to enhance the correlation physics, but not too strong to allow sufficiently small singlet-triplet gap.  
Recently, a curious magnetic behavior was observed in \syio* (SYIO), which was interpreted as excitonic magnetism.~\cite{cao14}
In particular, it was argued that non-cubic crystal field splits the triplet state and thus reduces the first 
excitation energy of the Ir ion. Subsequent band structure calculations~\cite{bhowal15} found only
moderate non-cubic crystal field on Ir site and found that the physics of SYIO is similar to its cubic
analog \byio* (BYIO). The GGA+U calculations to a magnetically ordered ground state, which the authors 
interpreted as confirmation of the experimental data.

In article we combine several methods to analyse the physics of BYIO and SYIO. 
Our results contradict the conclusions of Refs.~\onlinecite{bhowal15,cao14} as we find
BYIO and SYIO to be insulators with singlet local ground state.
We show that using the same GGA+U approach as in Ref.~\onlinecite{bhowal15} 
a solution corresponding to the singlet ground state has a lower energy than the solution
containing magnetic moments. This result is supported by
dynamical mean-field theory (DMFT) calculations performed with full interaction vertex.
We conclude with a sufficient margin for uncertainties of various model parameters
that the inter-atomic exchange processes in the studied double perovskites are too
weak to overcome the singlet-triplet splitting due to the spin-orbit coupling.
This result can be put to the perspective of the recent observation of signatures of 
excitonic magnetism in Ca$_2$RuO$_4$~\cite{CARUO}, a $4d$ perovskite where spin-orbit coupling is weaker
and Ru-Ru hopping is stronger than in Ir double perovskites.    

The paper is organized as follows. After introducing the computational methods in Sec.~\ref{sec:comp}
we discuss the bonding in double perovskite structure and construct an effective Hubbard model on the basis of 
band structure calculations (Sec.~\ref{sec:gga}). To compare with previous studies we have performed 
Hartree-like GGA+U calculations reported in Sec.~\ref{sec:ldau}. In the rest of the paper we analyze the multi-band
Hubbard model. First, we study the single-site problem with the focus on singlet-triplet splitting and
van Vleck susceptibility using exact diagonalization (Sec.~\ref{sec:ed}).
Next, we calculate the one-particle properties of the lattice problem using DMFT (Sec.~\ref{sec:dmft}) and
the dynamic of two-particle excitations in the strong-coupling expansion (Sec.~\ref{sec:t2u}).

\section{\label{sec:comp}Computational methods}
In this work we combine several numerical and semi-analytical approaches. All the below described
calculations were performed for cubic BYIO. The key steps were performed also for orthorhombic SYIO.

We start with GGA~\cite{gga} density functional calculations employing the Wien2k~\cite{wien2k} package.  
All the results reported here  were obtained with the spin-orbit coupling included~ \cite{singh94}. 
The crystallographic as well as other parameters used in the calculations are summarized in Supplemental Material (SM)~\cite{sm}. 

To include the correlation effects on the Ir site beyond the effective non-interacting picture of GGA we follow two paths. 
First, we use the static mean-field approach of GGA+U. Since this approach is known to often lead to multiple solutions 
we have run series of calculations for various initial conditions. In particular,
we used the spin-polarized GGA solution and atomic ground state as starting points.
The calculations were performed for interaction parameters $U$ of 2, 4 and 6~eV and $J$ from 0.1 to 0.7~eV.
The so called 'fully localized limit' double counting correction was used~\cite{FLL}. Although it should be a standard
for systems with strong spin-orbit coupling, 
we note explicitly that spin-off-diagonal terms of the '+U' orbital potential were included in the calculation. 

Second, we use the non-spin-polarized GGA band structures to construct effective Hubbard models.
Employing wien2wannier~\cite{w2w} and wannier90~\cite{wannier90} codes to construct the
Wannier representation of the GGA bands we can extract the crystal-field~\cite{novak13a} and spin-orbit
parameters as well as the inter-atomic hoppings. For the calculation of the partially screened Coulomb interaction,
we used the density response code for Elk~\cite{kozhevnikov10}.
In the calculation of the constrained susceptibility,
we excluded the contribution of particle-hole excitations within the
target bands (the $t_{2g}$ bands for the $d$-model, and  $t_{2g}$ and O $p$ bands
for the $pd$ model) and took 80 unoccupied bands.
We used 4$\times$4$\times$4 ${\bm k}$ and ${\bm q}$ meshes 
for the double Fourier transform with the cutoff $|{\bm G}+{\bm q}|$=1, 2, 3.5 (1/a.u.), 
where ${\bm G}$ is the reciprocal vector, and then extrapolated the result to 
$|{\bm G}+{\bm q}|=\infty$.

We study two types of model Hamiltonians: (i) $d$-model in the space of Ir-like $t_{2g}$ bands and
(ii) $pd$-model covering the space of Ir-$d$ and O-$p$ bands. The two models were expressed 
in two bases: (a) real cubic harmonics with sharp spin and (b) the eigenbasis of the local one-particle
Hamiltonian with \jo* doublet and \jt* quadruplet ($jj$-basis). The two bases are equivalent and 
their use followed numerical convenience.
We use both $d$- and $pd$-model primarily 
to demonstrate robustness of our results. The questions of $d$ vs $pd$ model
was discussed by several authors~\cite{} in the context of oxides of $3d$ elements. The basic difference
between the two models consists in how they describe low-energy one-particle excitations, whose orbital
character is a mixture of atomic-like Ir-$d$ and O-$p$ orbitals. In $d$-model these excitations are
molecular orbitals (anti-bonding $pd$ combinations) whose structure is determined solely by the $pd$ 
hybridization and does not depend on the interaction strength. In the $pd$-model the orbital structure 
of the excitations is affected by the interaction and can evolve from the non-interacting molecular orbital
limit to the Heitler-London limit where the double occupancy of the atomic orbitals is suppressed.
While the $pd$-model, which can describe richer physics, appears superior to the $d$-model, in practice
it is not necessarily so. Larger Hilbert space of the $pd$-model involves more uncertainties in its construction,
the most prominent of which is the so called double-counting correction which is directly related to the
$pd$ charge transfer energy. 

The basic information about the models is obtained by analyzing the local 
electronic structure. We performed exact diagonalization 
of the single-site $d$-model where the six $t_{2g}$ Wannier orbitals are treated as the atomic $d$ orbitals,
and iterative Lanczos diagonalization of the IrO${}_6$ cluster where the
complete Ir $d$ shell is hybridized with the O $p$ states. This step
served to determine the singlet-triplet splitting and to assess the
differences between the two models, mainly the consequences of the
explicit inclusion of the $pd$ hybridization in the cluster model.

To extend the many-body calculations on the lattice we have run DMFT simulations for both the $d$ and the $pd$ model
with the interaction parameters obtained in the cRPA calculations. To this end we used our 
implementation of the self-consistent cycle~\cite{kunes07} and the TRIQS hybridization-expansion 
continuous-time quantum Monte Carlo solver~\cite{sk2015} (based on the TRIQS library~\cite{pf2015}).
The calculations, 
with full Coulomb vertex $U_{ijkl}c_i^\dag c_j^\dag c_l^{\phantom{\dag}} c_k^{\phantom{\dag}}$,
were performed in \jo*-\jt* basis ($jj$-basis) in which the on-site one-particle Hamiltonian is diagonal.
This choice  of basis helps to avoid the Monte Carlo sign problem.~\cite{ss2015}
The reported calculations were performed at the temperature of 290~K.
The spectral functions were obtained by analytic continuation of the self-energy using the maximum entropy method~\cite{MEM}.

While our DMFT calculations were limited to studying one-particle dynamics, we use large-$U$ expansion of the $d$-model to 
investigate the two-particle excitations. We have eliminated the charge fluctuations by the Schrieffer-Wolff transformation~\cite{sw}
to the second order and constructed an effective Hamiltonian on a local Hilbert space spanned by the singlet and lowest
triplet states. The dynamics of this effective model was studied in the linear spin wave approximation.
To check the consistency of our results we have started from two different one-particle bases: the real cubic harmonics
and $jj$-basis. 

\section{Results and discussion}
\subsection{\label{sec:gga}GGA}
First, we report the unpolarized GGA solution for BYIO, which served as a starting point to construct the effective
models. The results for lower-symmetry SYIO are similar.
\subsubsection{Molecular orbitals}
Unlike in simple perovskites, where the O$_6$ octahedra around neighboring metal atoms share corners, in double perovskites
the IrO$_6$ octahedra are not touching. This geometry suggests that the double perovskites can be possibly viewed
as molecular crystals built out of IrO$_6$ molecules plus some filler atoms. Irrespective of whether this picture
is useful for interpretation of the physics, one can use molecular orbitals (MOs) - linear combinations of atom-centered Wannier functions (WFs) on the O$_6$ 
octahedron - as an alternative to the atom-centered WF basis.

\begin{figure}
\includegraphics[width=\columnwidth,clip]{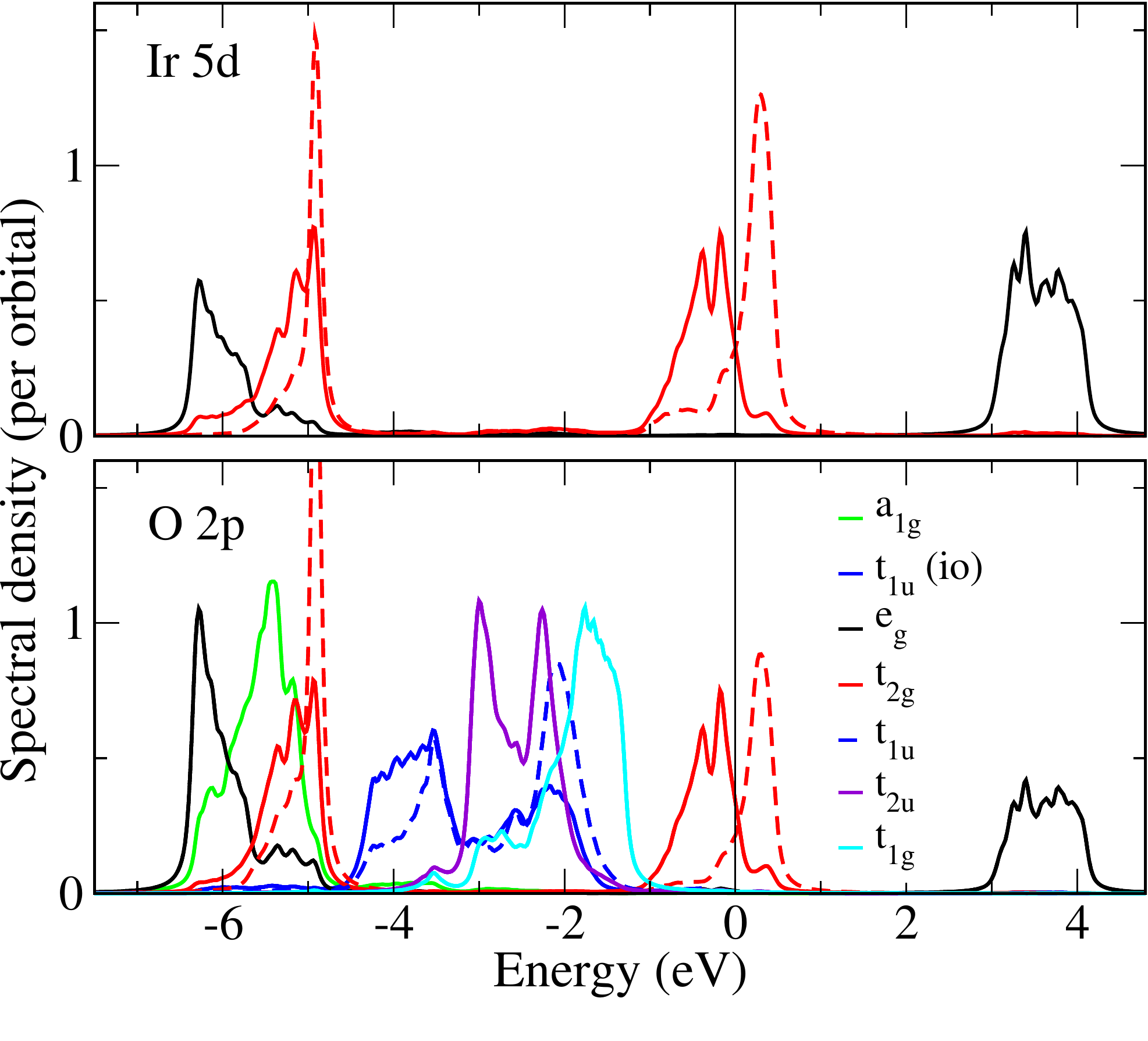}
\caption{\label{fig:mo} The GGA+so spectral density of BYIO projected on the irreducible representations
of the cubic symmetry in the space of $pd$ Wannier functions. Top panel: projection on WFs centered on the Ir atom
(the $t_{2g}$ functions were further resolved into the \jt* and \jo* states). Bottom panel: 
projection on WFs centered on the O atoms. The $a_{1g}$, $e_g$ and $t_{1u} (io)$ MOs
are built from atomic O-$p$ orbitals pointing to the central Ir atom, while the remaining MOs are built from 
atomic O-$p$ orbitals perpendicular to the Ir-O bond.}
\end{figure}

\begin{figure}
\includegraphics[width=\columnwidth,clip]{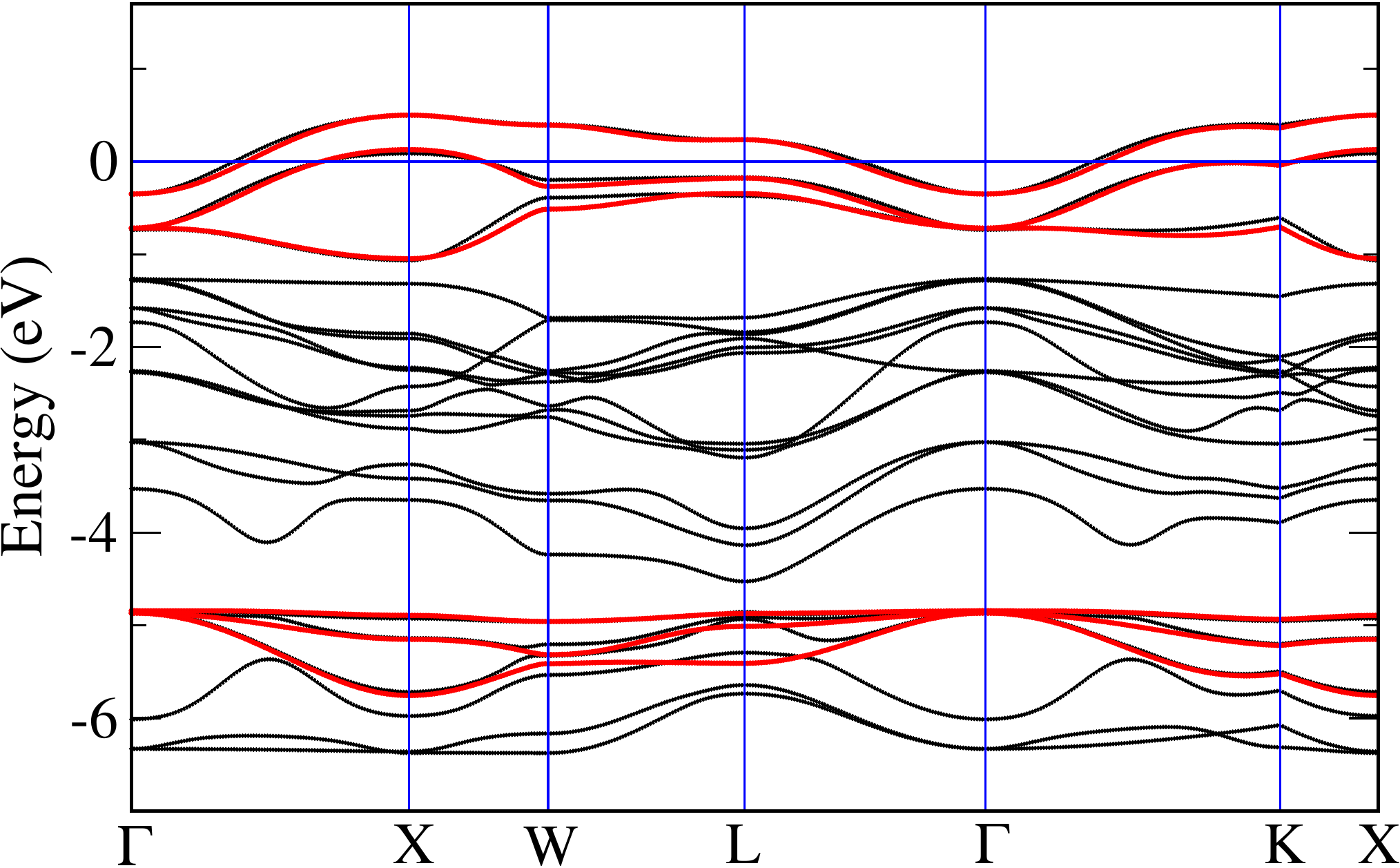}
\caption{\label{fig:bands} The GGA+so band structure of BYIO (black). The bands
of the $pd$ model with $t_{2g}$ orbitals described in the text (red).}
\end{figure}

In Fig.~\ref{fig:mo} we show the GGA spectral density decomposed into the projections on the MOs classified
by their symmetry. 
First, we note the magnitude of the $pd$ hybridization. Nominally Ir-$d$ orbitals
around Fermi level are superpositions of the atomic Ir-$d$ orbital and its MO partner. For \jt* states
the mixture is essentially fifty-fifty, while for \jo* and $e_g$ states the Ir-$d$ weight is somewhat larger.
We further observe that bands of distinct symmetry exhibit little hybridization.
The filled non-bonding MOs therefore do not play an active role in the low-energy physics.
In Fig.~\ref{fig:bands} it is demonstrated that they can be dropped from the model.
We compare the GGA band structure with the bands of a truncated $pd$ model where only Ir and O$_6$ $t_{2g}$ 
orbitals were kept, while the $e_g$ and non-bonding orbitals were simply left out without 
any adjustment of the $t_{2g}$ subspace. 
The model dispersion reproduces very well the anti-bonding bands around the Fermi level and
does a fairly good job also for the bonding bands which overlap with the $e_g$ and $a_{1g}$ bands.
This suggests that the molecular crystal provides a useful picture of the studied double perovskites.
The large bonding--anti-bonding splitting allows straightforward construction of the Wannier representation
of the anti-bonding bands only -  the $d$ model.

\subsubsection{cRPA}
The cRPA calculations were performed for $d$ and $pd$ models. The calculated direct and exchange
interaction parameters in the low frequency limit can be found in SM. In our model calculations we make the common
assumption of atomic form of the interaction, which is then parametrized by $U$ and $J$ which are chosen
to match the cRPA data. For the $d$-model we can fairly accurately fit the cRPA data 
with the values $U$=1.8~eV and $J$=0.4~eV, while we get $U$=4~eV and $J$=1.1~eV for the $pd$-model. 
The values of $J$ should be viewed as upper estimates obtained by the extrapolation of the Fourier 
cut-off to infinity. Our main conclusions in this work hold for smaller  $J$, but eventually break for large $J$. 
Therefore we try to use conservative (upper limit) values in our calculations.

\subsubsection{\label{sec:ldau}Spin-polarised solution and GGA+U}
\begin{figure}
\includegraphics[width=0.8\columnwidth,clip]{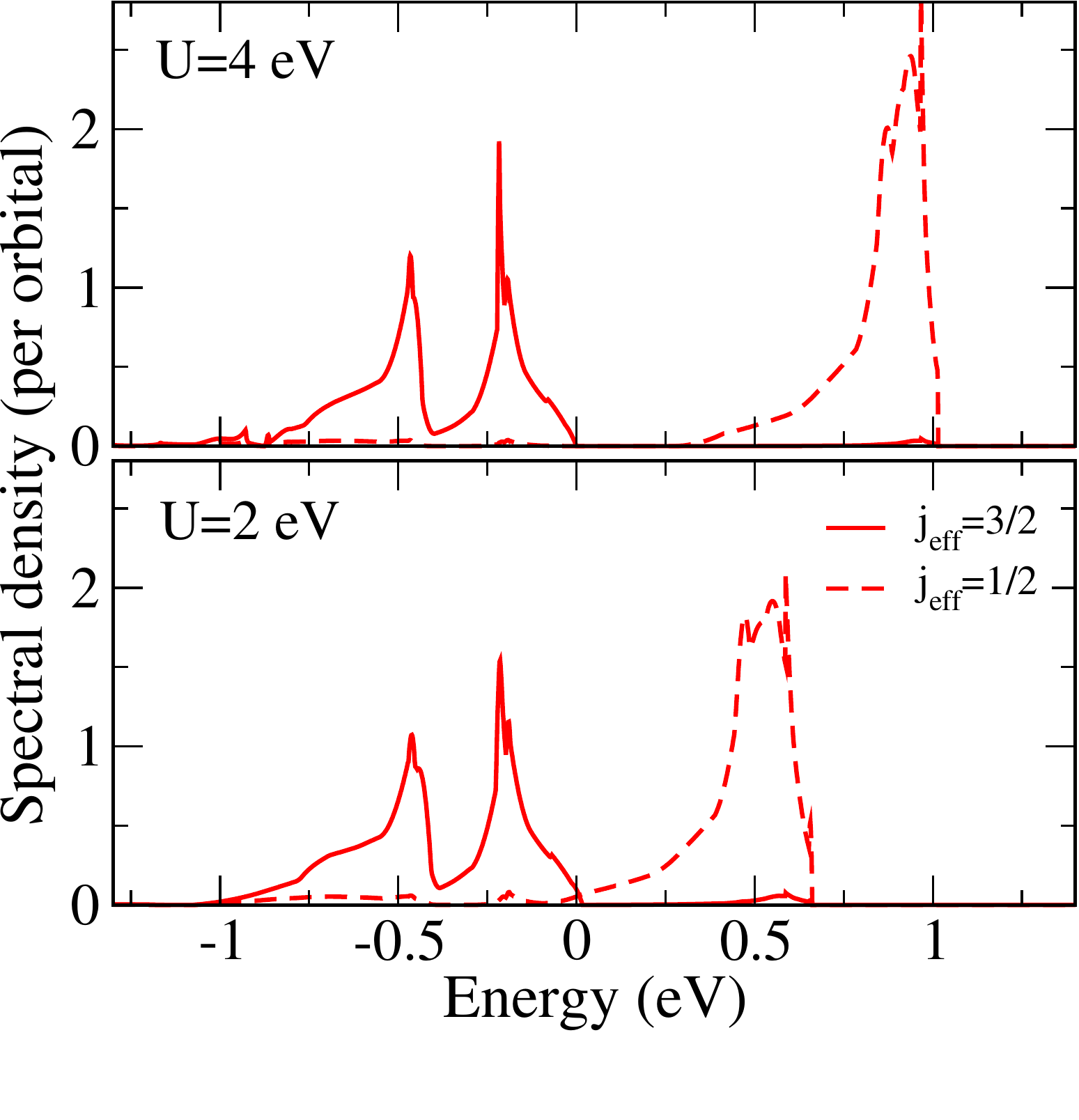}
\caption{\label{fig:ldau} The Ir $t_{2g}$ spectral densities for the non-magnetic GGA+U solutions obtained with $J=0.6$~eV and 
two values of $U$.}
\end{figure}

Allowing ferromagnetic spin-polarized solutions we arrive at similar results as reported in Ref.~\onlinecite{bhowal15}. However,
these solutions are metallic, contrary to experiment, suggesting that inclusion of electronic correlations beyond 
GGA is desirable. The GGA+U approach provides a computationally cheap way to include the electronic correlations
on a Hatree-Fock level. The GGA+U method is formally analogous to the Weiss mean-field theory and as such it is prone to 
multiple self-consistent solutions that could be thermodynamically unstable. 
The authors of Ref.~\onlinecite{bhowal15} started from their spin polarized solutions for BYIO and SYIO  and including 
the +U orbital potential they found these solutions to be modified, but qualitatively unchanged. While we were able to 
find similar spin-polarized solutions (SM), we extended the search for possible stable solutions by starting from
the occupation matrices corresponding to the atomic ground state reported below. This way we obtained non-magnetic solutions 
with energies lower than their spin polarized counterparts. The spectral density of the non-magnetic solution for 
BYIO, shown in Fig.~\ref{fig:ldau}, confirms the picture of a local singlet ground state with filled \jt* orbitals and empty \jo* ones. 

\subsection{Multi-orbital Hubbard model}
In the following, we present the results obtained for the $d$ and $pd$ models of SYIO and BYIO. The parameters
of the models obtained using the Wannier projection can be found in SM.

\subsubsection{\label{sec:ed}Single-site physics}
\begin{figure}[!t]
\begin{tabular}[b]{lr}
\includegraphics[width=0.43\columnwidth,clip]{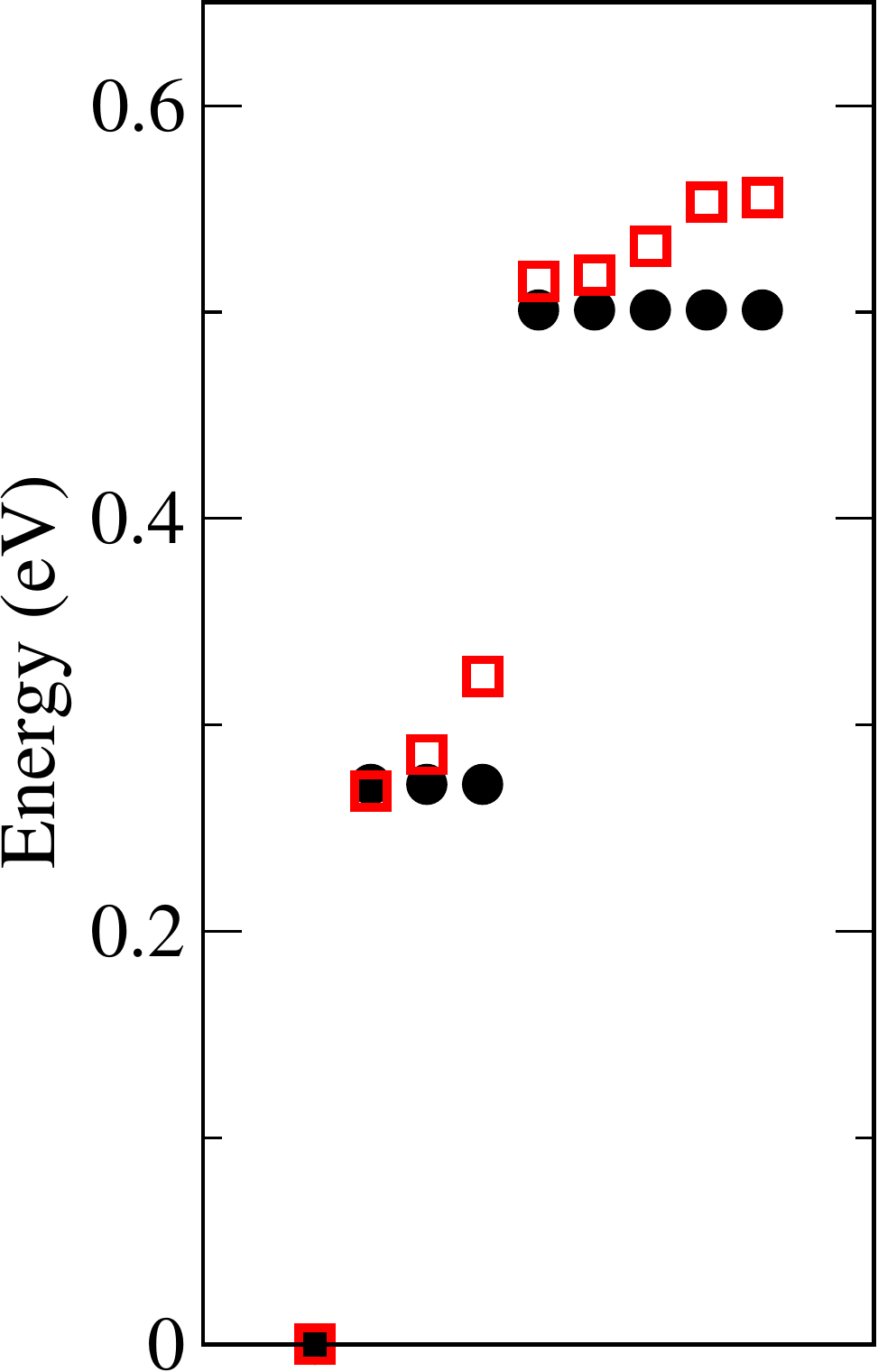} &
\includegraphics[width=0.45\columnwidth,clip]{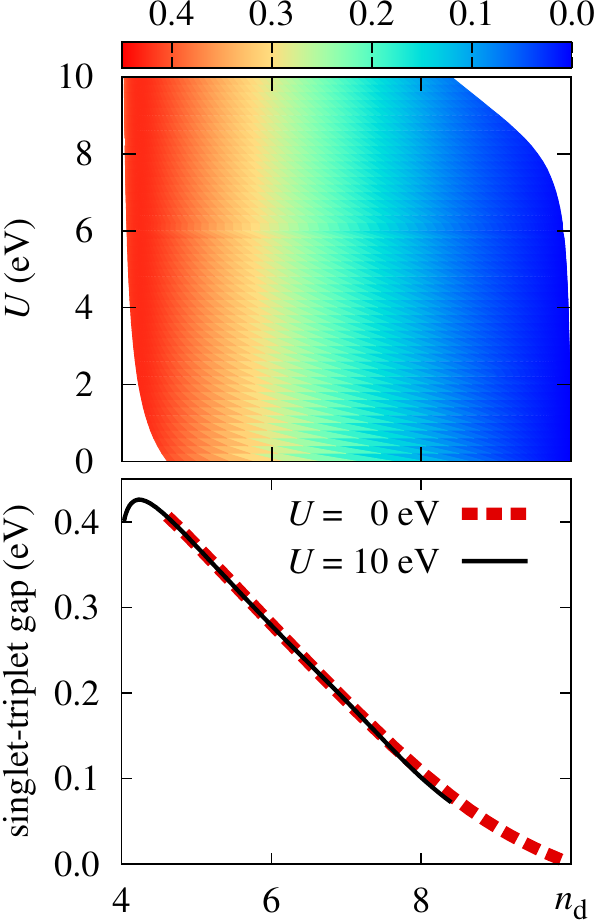}\\
\end{tabular}
\caption{\label{fig:spec} (a) The lowest local atomic multiplets of the $d^4$ state
in BYIO (black circles) and SYIO (red squares) for ${J=0.4}$~eV and $d$-model.
(b) The singlet-triplet splitting in the IrO$_6$ cluster ($pd$-model) for $J=1.1$~eV
as a function of $U$ and Ir-$d$ occupancy. (c) Horizontal cuts in (b) at $U=0$ and $U=10$~eV.}
\end{figure}

We start by discussing the physics of an isolated lattice site. The key quantity of interest is the singlet-triplet excitation energy $\Delta_t$.  
In Fig.~\ref{fig:spec} we show the single-ion spectra of BYIO and SYIO  of the $d$-model
obtained by diagonalization of the single-site Hamiltonian with 4 electrons.  
Similar to Ref.~\onlinecite{bhowal15} we find that the non-cubic crystal-field
in SYIO has only minor effect and the spectrum is controlled by the spin-orbit coupling and Hund's 
exchange $J$  (Note that the isotropic part $U$ of the on-site interaction does not affect level splitting 
within a given valence state.) Since the singlet-triplet splitting decreases with $J$, 
the excitation energy $\Delta_t$ of 270~meV is a conservative lower estimate.

Besides the multiplet structure we have calculated the van Vleck susceptibility of the Ir atom in the $d$-model~\cite{novak13b} of SYIO.
The susceptibility increases with $J$ and it exhibits an anisotropy: for $J$=0.4~eV its values along a, b and c axes are
0.0011, 0.0016 and 0.0009 emu/mol (the corresponding values for $J$=0.2~eV are 0.0009, 0.0009 and 0.0008
emu/mole). These numbers correspond rather well to the experimental data of Ref.~\onlinecite{cao14} which exhibit
only a weak temperature dependence above 50~K ($\sim$0.0015~emu/mole@50~K and $\sim$0.0011~emu/mole@330~K).

Next , we study the local physics of BYIO within the $pd$-model, i.e. we diagonalize the IrO$_6$ cluster. We treat the O-$p$ 
states as non-interacting and put the interaction on the central Ir atom only. 
Unlike for the isolated $d$-shell above, the isotropic part of the interaction $U$  affects the results strongly. 
The dependence of the excitation spectrum on $U$ is, however, to a large extent compensated
by the double-counting correction $E_{\text{dc}}$ which renormalizes the splitting between
the $d$ and $p$ orbital energies and which must be included to take into account
the interaction energy present in GGA.
Since $E_{\text{dc}}$ is not rigorously defined, we treat it as an adjustable parameter.
In Fig.~\ref{fig:spec} we show the singlet-triplet excitation energy $\Delta_t$ for the IrO$_6$ cluster.
Rather than $E_{\text{dc}}$, we plot our results as a function of the closely related Ir-$d$ occupancy $n_d$ with
clear physical meaning. For fixed $n_d$ the excitation energy $\Delta_t$ is practically independent of $U$.
We take the Ir-$d$ occupancies between $6-7$ ($n_d=6.35$  is the GGA value) to be physically reasonable. 
The corresponding singlet-triplet splitting falls into the interval 190-280 meV and is consistent with 270~meV 
obtained from the $d$-model.  
 
As a side remark we point out that  a popular approximation of the interaction to the density-density terms in the $jj$-basis leads
to the substantially smaller values of singlet-triplet splitting of around 100~meV. This can be traced back to the missing
pair-hopping terms which contribute to lowering of the singlet energy.

\subsubsection{\label{sec:dmft}DMFT}
In the following, we extend our analysis to the lattice problem. We address three questions. Is there a charge gap in the 
one-particle spectrum and what is its nature? It there a global ground state with broken symmetry predicted by Ref.~\onlinecite{bhowal15}?
What is the dynamics of two-particle excitations and can they condense in the sense proposed by \kha*~\cite{kha13}?
The first and, to some extent, the second questions are studied using DMFT. The second and the third questions are addressed using the effective strong coupling
Hamiltonian.

\begin{figure}
\includegraphics[width=\columnwidth,clip]{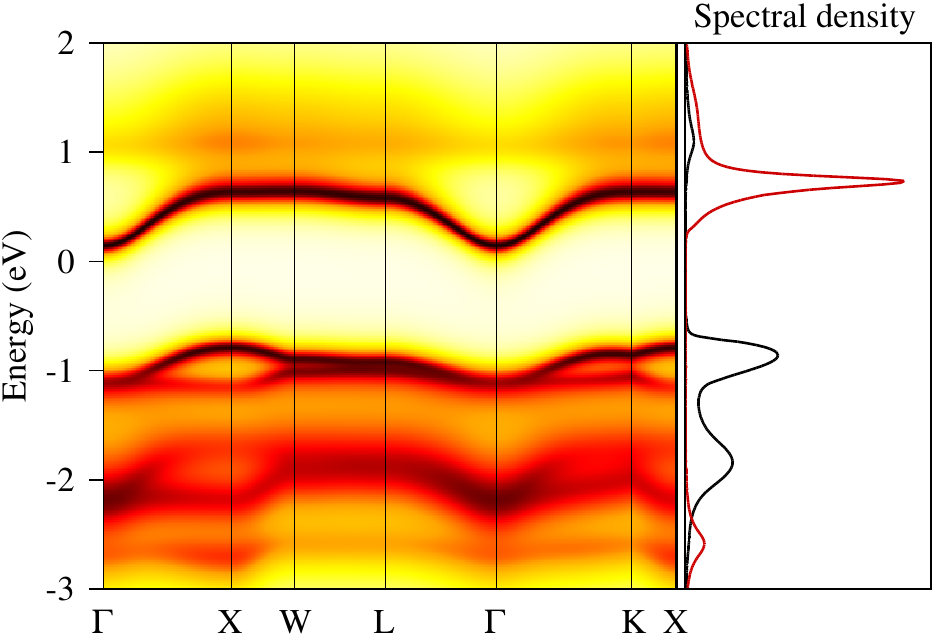}
\caption{\label{fig:dmft} The one-particle spectral density for the $d$-model of BYIO along high symmetry directions
in the Brillouin zone obtained with DMFT. The side panel shows the k-integrated density resolved
into \jo* (red) and \jt* (black) orbital contributions. The calculation were performed with $U=$1.8~eV, $J=$0.4~eV 
at the temperature of 290~K.}
\end{figure}

The DMFT technique incorporates local quantum and thermal fluctuations and does not typically suffer from the problem of multiple
solutions with distinct local states. When several local states are quasi-degenerate the method can naturally describe
their statistical mixture (most common example being local moment fluctuations) or even their condensate~\cite{kunes14a,kunes14b}. Moreover,
modern DMFT implementations can treat full Coulomb vertex without approximations. In Fig.~\ref{fig:dmft} we show
the DMFT one-particle spectra of the $d$ model.  
The dominant correlation effect is the static Hartree shift of the empty band with dominant \jo* character with respect
to the occupied band dominated by the \jt* character. The dynamical effects, arising from the pair-hopping terms in the
interaction play only a minor role. The picture provided by DMFT is therefore consistent with the GGA+U non-magnetic solution. 

\subsubsection{\label{sec:t2u}Strong coupling expansion}
\begin{figure}
\includegraphics[width=\columnwidth,clip]{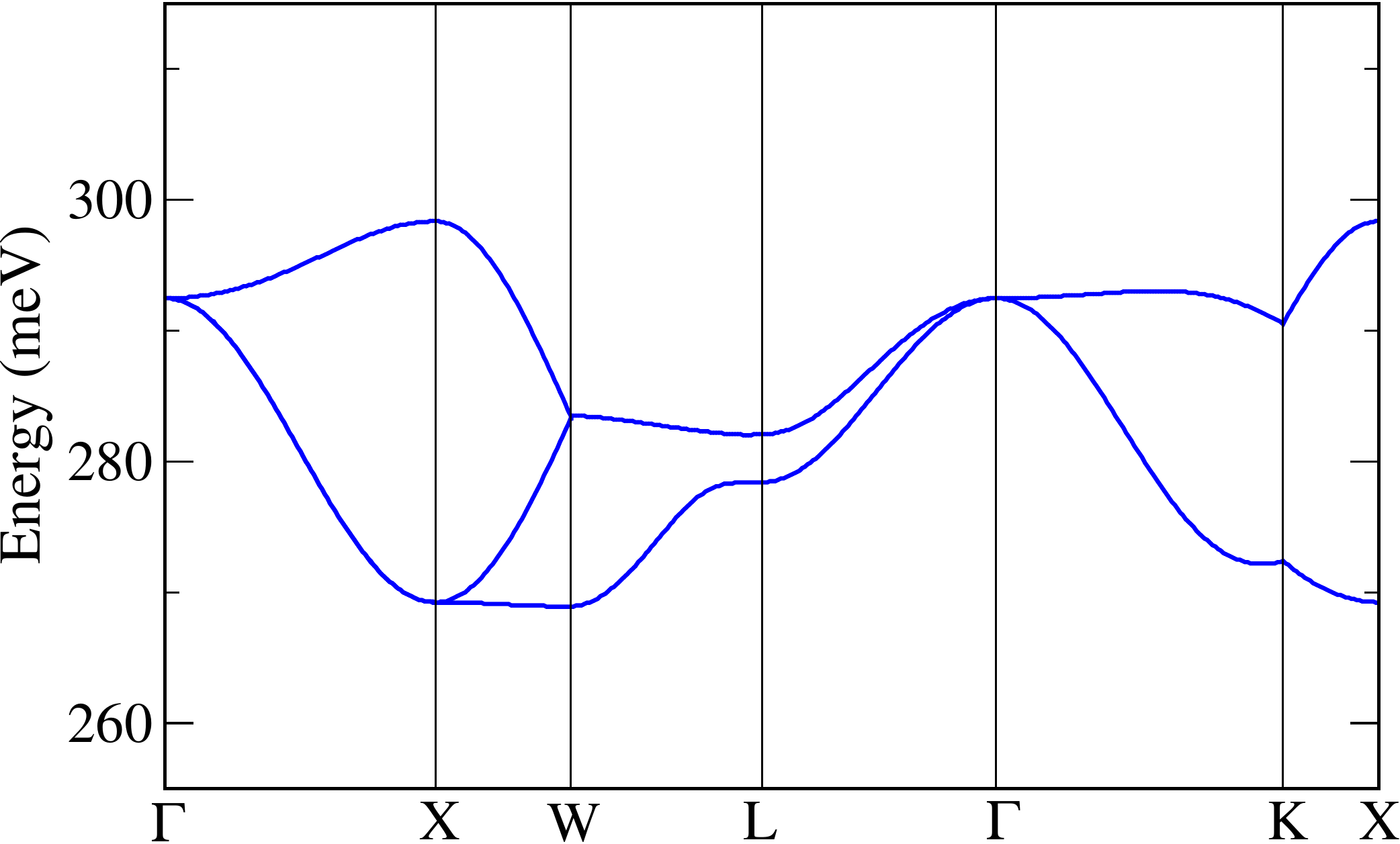}
\caption{\label{fig:lsw} Dispersion of a triplet excitation in singlet background obtained from $d$-model with $U=1.8$~eV and ${J=0.4}$~eV.}
\end{figure}

In order to assess the possibility of excitonic condensation we have to investigate the two-particle dynamics.
While it can, in principle, be studied by DMFT, in practice this is a difficult task which has been executed only for simple
models or involving approximation.~\cite{park11,boehnke12} 
Moreover, the results so far suggest that BYIO (and SYIO) can be viewed as strongly coupled systems with suppressed charge
fluctuations. An effective strong coupling model provides a much more accessible description of the physics.
In the following, we derive such a model and solve its dynamics using the linear spin-wave theory.
The model in the space spanned by the singlet \s* and triplet \t* states 
is obtained by second-order Schrieffer-Wolff transformation of the $d$ model. 
The classical ground state is a product of singlets. The excitations in the limit of low \t*-density 
are described by an effective Hamiltonian describing propagation of a single $t$-excitation, i.e., we neglect for example interactions between $t$s,
\begin{equation}
\label{eq:spw}
\begin{split}
&H=\varepsilon \sum_{i,s}  t_{is}^{\dagger}t_{is}{\phantom\dagger}\\
&+\frac{1}{2}\sum_{i,\bR,s,s'}\left(\Lambda_{ss'}^{\bR}t^{\dagger}_{is} t^{\phantom\dagger}_{i+{\bR}s'}
+\Gamma_{ss'}^{\bR}t^{\dagger}_{is}t^{\dagger}_{i+{\bR}s'}\right)+H.c.
\end{split}
\end{equation}
Here, $\varepsilon$ is the singlet-triplet splitting $\Delta_t$ renormalised by hopping processes of the type $|s,s\rangle\rightarrow|s,s\rangle$ and 
$|s,t_s\rangle\rightarrow|s,t_s\rangle$, $\Lambda_{ss'}^{\bR}$ is the amplitude of the process $|s,t_s\rangle\rightarrow|t_{s'},s\rangle$ and
$\Gamma_{ss'}^{\bR}$ is the amplitude of the process $|s,s\rangle\rightarrow|t_{s'},t_s\rangle$. Hopping to the first and second neighbors
on the fcc lattice is considered. In Fig.~\ref{fig:lsw} we present the spectrum of (\ref{eq:spw}) along the high symmetry directions of the cubic
lattice. The main observation is the large excitation gap. In the language of the \kha*'s proposal the excitonic condensation 
takes place when the energy of the lowest triplet excitation touches the singlet (vacuum) level. Our results show that the width of the $t$-dispersion
is at least an order of magnitude too small for that to happen. Since the parameters of (\ref{eq:spw}) for SYIO are of similar magnitude as for BYIO 
we conclude that our result holds for SYIO as well.    

The double perovskite structure with large Ir-Ir distance and small hopping amplitude $t$ is not very promising for observation of excitonic magnetism. 
The electronic bandwidth $W\sim Zt$ does not have to be unusually small thanks to the high number of nearest neighbors $Z$.
The effect of small $t$ is much more pronounced in the dispersion of two-particle excitations with a bandwidth $W_t\sim Z t^2$ which
scales as $1/Z$ for fixed $W$.  

\section{Conclusions}
We have combined several theoretical approaches to study the low-energy properties of Ir double perovskites BYIO and SYIO.
The two compounds are quite similar, in particular, the non-cubic crystal field of SYIO is found to be rather weak, a result that agrees
with the conclusions of theoretical study of Bhowal {\it et al.}~\cite{bhowal15}. Our main conclusion is that excitonic magnetism 
in BYIO and SYIO is not favored. Small Ir-Ir hopping in double perovskite structure gives rise to valence and conduction bands that
are sufficiently narrow to become orbitally polarized and gapped by a moderate on-site Coulomb repulsion. Another consequence of the 
small electron hopping are small amplitudes for propagation of two-particle excitations. The bandwidth of such excitations is an order 
of magnitude smaller than the singlet-triplet excitation energy determined by the spin-orbit coupling and Hund's exchange $J$, which prevents the possibility 
of their exciton condensation.

Our conclusions contradict the conclusions of the experimental study~\cite{cao14}. While our calculated van Vleck susceptibility 
reproduces quite well the experimental susceptibility of Ref.~\onlinecite{cao14} above 50~K, we do not find any tendency towards
a magnetic phase transition at low temperatures. Instead, we predict a large spin gap in the 200 meV range. 
Our results thus call for reexamination of the experimental data considering the possibility of an extrinsic origin of the observed low temperature behavior.

\acknowledgements
We acknowledge discussions with Z. Jir\'ak, K. Kn\'{i}\v{z}ek and J. Hejtm\'anek.  We acknowledge support of the Czech Science Foundation
through the project 13-25251S (JKu and PN), German Research Foundation through FOR1346 (VP) and Grants-in-Aids for Scientific Research on Innovative Areas (15H05883) from JSPS (RA). Access to computing and storage facilities owned by parties and projects contributing to the National Grid Infrastructure MetaCentrum, provided under the programme "Projects of Large Infrastructure for Research, Development, and Innovations" (LM2010005), is greatly appreciated.


\begin{thebibliography}{25}
\expandafter\ifx\csname natexlab\endcsname\relax\def\natexlab#1{#1}\fi
\expandafter\ifx\csname bibnamefont\endcsname\relax
  \def\bibnamefont#1{#1}\fi
\expandafter\ifx\csname bibfnamefont\endcsname\relax
  \def\bibfnamefont#1{#1}\fi
\expandafter\ifx\csname citenamefont\endcsname\relax
  \def\citenamefont#1{#1}\fi
\expandafter\ifx\csname url\endcsname\relax
  \def\url#1{\texttt{#1}}\fi
\expandafter\ifx\csname urlprefix\endcsname\relax\def\urlprefix{URL }\fi
\providecommand{\bibinfo}[2]{#2}
\providecommand{\eprint}[2][]{\url{#2}}

\bibitem[{\citenamefont{Kim et~al.}(2008)\citenamefont{Kim, Jin, Moon, Kim,
  Park, Leem, Yu, Noh, Kim, Oh et~al.}}]{kim08}
\bibinfo{author}{\bibfnamefont{B.~J.} \bibnamefont{Kim}},
  \bibinfo{author}{\bibfnamefont{H.}~\bibnamefont{Jin}},
  \bibinfo{author}{\bibfnamefont{S.~J.} \bibnamefont{Moon}},
  \bibinfo{author}{\bibfnamefont{J.-Y.} \bibnamefont{Kim}},
  \bibinfo{author}{\bibfnamefont{B.-G.} \bibnamefont{Park}},
  \bibinfo{author}{\bibfnamefont{C.~S.} \bibnamefont{Leem}},
  \bibinfo{author}{\bibfnamefont{J.}~\bibnamefont{Yu}},
  \bibinfo{author}{\bibfnamefont{T.~W.} \bibnamefont{Noh}},
  \bibinfo{author}{\bibfnamefont{C.}~\bibnamefont{Kim}},
  \bibinfo{author}{\bibfnamefont{S.-J.} \bibnamefont{Oh}},
  \bibnamefont{et~al.}, \bibinfo{journal}{Phys. Rev. Lett.}
  \textbf{\bibinfo{volume}{101}}, \bibinfo{pages}{076402}
  (\bibinfo{year}{2008}).

\bibitem[{\citenamefont{Khaliullin}(2013)}]{kha13}
\bibinfo{author}{\bibfnamefont{G.}~\bibnamefont{Khaliullin}},
  \bibinfo{journal}{Phys. Rev. Lett.} \textbf{\bibinfo{volume}{111}},
  \bibinfo{pages}{197201} (\bibinfo{year}{2013}).

\bibitem[{\citenamefont{Cao et~al.}(2014)\citenamefont{Cao, Qi, Li, Terzic,
  Yuan, DeLong, Murthy, and Kaul}}]{cao14}
\bibinfo{author}{\bibfnamefont{G.}~\bibnamefont{Cao}},
  \bibinfo{author}{\bibfnamefont{T.~F.} \bibnamefont{Qi}},
  \bibinfo{author}{\bibfnamefont{L.}~\bibnamefont{Li}},
  \bibinfo{author}{\bibfnamefont{J.}~\bibnamefont{Terzic}},
  \bibinfo{author}{\bibfnamefont{S.~J.} \bibnamefont{Yuan}},
  \bibinfo{author}{\bibfnamefont{L.~E.} \bibnamefont{DeLong}},
  \bibinfo{author}{\bibfnamefont{G.}~\bibnamefont{Murthy}}, \bibnamefont{and}
  \bibinfo{author}{\bibfnamefont{R.~K.} \bibnamefont{Kaul}},
  \bibinfo{journal}{Phys. Rev. Lett.} \textbf{\bibinfo{volume}{112}},
  \bibinfo{pages}{056402} (\bibinfo{year}{2014}).

\bibitem[{\citenamefont{Bhowal et~al.}(2015)\citenamefont{Bhowal, Baidya,
  Dasgupta, and Saha-Dasgupta}}]{bhowal15}
\bibinfo{author}{\bibfnamefont{S.}~\bibnamefont{Bhowal}},
  \bibinfo{author}{\bibfnamefont{S.}~\bibnamefont{Baidya}},
  \bibinfo{author}{\bibfnamefont{I.}~\bibnamefont{Dasgupta}}, \bibnamefont{and}
  \bibinfo{author}{\bibfnamefont{T.}~\bibnamefont{Saha-Dasgupta}},
  \bibinfo{journal}{Phys. Rev. B} \textbf{\bibinfo{volume}{92}},
  \bibinfo{pages}{121113} (\bibinfo{year}{2015}).

\bibitem[{\citenamefont{Jain et~al.}()\citenamefont{Jain, Krautloher, Porras,
  Ryu, Chen, Abernathy, Park, Ivanov, Chaloupka, Khaliullin et~al.}}]{CARUO}
\bibinfo{author}{\bibfnamefont{A.}~\bibnamefont{Jain}},
  \bibinfo{author}{\bibfnamefont{M.}~\bibnamefont{Krautloher}},
  \bibinfo{author}{\bibfnamefont{J.}~\bibnamefont{Porras}},
  \bibinfo{author}{\bibfnamefont{G.~H.} \bibnamefont{Ryu}},
  \bibinfo{author}{\bibfnamefont{D.~P.} \bibnamefont{Chen}},
  \bibinfo{author}{\bibfnamefont{D.~L.} \bibnamefont{Abernathy}},
  \bibinfo{author}{\bibfnamefont{J.~T.} \bibnamefont{Park}},
  \bibinfo{author}{\bibfnamefont{A.}~\bibnamefont{Ivanov}},
  \bibinfo{author}{\bibfnamefont{J.}~\bibnamefont{Chaloupka}},
  \bibinfo{author}{\bibfnamefont{G.}~\bibnamefont{Khaliullin}},
  \bibnamefont{et~al.}, \bibinfo{note}{arXiv:1510.07011}.

\bibitem[{\citenamefont{Perdew et~al.}(1996)\citenamefont{Perdew, Burke, and
  Ernzerhof}}]{gga}
\bibinfo{author}{\bibfnamefont{J.~P.} \bibnamefont{Perdew}},
  \bibinfo{author}{\bibfnamefont{K.}~\bibnamefont{Burke}}, \bibnamefont{and}
  \bibinfo{author}{\bibfnamefont{M.}~\bibnamefont{Ernzerhof}},
  \bibinfo{journal}{Phys. Rev. Lett.} \textbf{\bibinfo{volume}{77}},
  \bibinfo{pages}{3865} (\bibinfo{year}{1996}).

\bibitem[{\citenamefont{Blaha et~al.}(2001)\citenamefont{Blaha, Schwarz,
  Madsen, Kvasnicka, and Luitz}}]{wien2k}
\bibinfo{author}{\bibfnamefont{P.}~\bibnamefont{Blaha}},
  \bibinfo{author}{\bibfnamefont{K.}~\bibnamefont{Schwarz}},
  \bibinfo{author}{\bibfnamefont{G.~K.~H.} \bibnamefont{Madsen}},
  \bibinfo{author}{\bibfnamefont{D.}~\bibnamefont{Kvasnicka}},
  \bibnamefont{and} \bibinfo{author}{\bibfnamefont{J.}~\bibnamefont{Luitz}},
  \emph{\bibinfo{title}{WIEN2k, An Augmented Plane Wave + Local Orbitals
  Program for Calculating Crystal Properties}} (\bibinfo{publisher}{Technische
  Universit\"{a}t Wien}, \bibinfo{address}{Austria}, \bibinfo{year}{2001}).

\bibitem[{\citenamefont{Singh}(1994)}]{singh94}
\bibinfo{author}{\bibfnamefont{D.~J.} \bibnamefont{Singh}},
  \emph{\bibinfo{title}{Planewaves, Pseudopotentials and the LAPW Method}}
  (\bibinfo{publisher}{Kluwer Academic}, \bibinfo{address}{Dordrecht},
  \bibinfo{year}{1994}).

\bibitem[{sm()}]{sm}
\bibinfo{note}{See Supplemental Material},
  \urlprefix\url{http://www.fzu.cz/~kunes/papers/BYIO_sm.pdf}.

\bibitem[{\citenamefont{Liechtenstein et~al.}(1995)\citenamefont{Liechtenstein,
  Anisimov, and Zaanen}}]{FLL}
\bibinfo{author}{\bibfnamefont{A.~I.} \bibnamefont{Liechtenstein}},
  \bibinfo{author}{\bibfnamefont{V.~I.} \bibnamefont{Anisimov}},
  \bibnamefont{and} \bibinfo{author}{\bibfnamefont{J.}~\bibnamefont{Zaanen}},
  \bibinfo{journal}{Phys. Rev. B} \textbf{\bibinfo{volume}{52}},
  \bibinfo{pages}{R5467} (\bibinfo{year}{1995}).

\bibitem[{\citenamefont{Kune\v{s} et~al.}(2010)\citenamefont{Kune\v{s}, Arita,
  Wissgott, Toschi, Ikeda, and Held}}]{w2w}
\bibinfo{author}{\bibfnamefont{J.}~\bibnamefont{Kune\v{s}}},
  \bibinfo{author}{\bibfnamefont{R.}~\bibnamefont{Arita}},
  \bibinfo{author}{\bibfnamefont{P.}~\bibnamefont{Wissgott}},
  \bibinfo{author}{\bibfnamefont{A.}~\bibnamefont{Toschi}},
  \bibinfo{author}{\bibfnamefont{H.}~\bibnamefont{Ikeda}}, \bibnamefont{and}
  \bibinfo{author}{\bibfnamefont{K.}~\bibnamefont{Held}},
  \bibinfo{journal}{Computer Physics Communications}
  \textbf{\bibinfo{volume}{181}}, \bibinfo{pages}{1888 }
  (\bibinfo{year}{2010}).

\bibitem[{\citenamefont{Mostofi et~al.}(2008)\citenamefont{Mostofi, Yates, Lee,
  Souza, Vanderbilt, and Marzari}}]{wannier90}
\bibinfo{author}{\bibfnamefont{A.~A.} \bibnamefont{Mostofi}},
  \bibinfo{author}{\bibfnamefont{J.~R.} \bibnamefont{Yates}},
  \bibinfo{author}{\bibfnamefont{Y.-S.} \bibnamefont{Lee}},
  \bibinfo{author}{\bibfnamefont{I.}~\bibnamefont{Souza}},
  \bibinfo{author}{\bibfnamefont{D.}~\bibnamefont{Vanderbilt}},
  \bibnamefont{and} \bibinfo{author}{\bibfnamefont{N.}~\bibnamefont{Marzari}},
  \bibinfo{journal}{Computer Physics Communications}
  \textbf{\bibinfo{volume}{178}}, \bibinfo{pages}{685 } (\bibinfo{year}{2008}).

\bibitem[{\citenamefont{Nov\'ak
  et~al.}(2013{\natexlab{a}})\citenamefont{Nov\'ak,
  Kn\'{\i}\ifmmode~\check{z}\else \v{z}\fi{}ek, and Kune\ifmmode~\check{s}\else
  \v{s}\fi{}}}]{novak13a}
\bibinfo{author}{\bibfnamefont{P.}~\bibnamefont{Nov\'ak}},
  \bibinfo{author}{\bibfnamefont{K.}~\bibnamefont{Kn\'{\i}\ifmmode~\check{z}\else
  \v{z}\fi{}ek}}, \bibnamefont{and}
  \bibinfo{author}{\bibfnamefont{J.}~\bibnamefont{Kune\ifmmode~\check{s}\else
  \v{s}\fi{}}}, \bibinfo{journal}{Phys. Rev. B} \textbf{\bibinfo{volume}{87}},
  \bibinfo{pages}{205139} (\bibinfo{year}{2013}{\natexlab{a}}).

\bibitem[{\citenamefont{Kozhevnikov et~al.}(2010)\citenamefont{Kozhevnikov,
  Eguiluz, and Schulthess}}]{kozhevnikov10}
\bibinfo{author}{\bibfnamefont{A.}~\bibnamefont{Kozhevnikov}},
  \bibinfo{author}{\bibfnamefont{A.}~\bibnamefont{Eguiluz}}, \bibnamefont{and}
  \bibinfo{author}{\bibfnamefont{T.}~\bibnamefont{Schulthess}}, in
  \emph{\bibinfo{booktitle}{SC10: Proceedings of the 2010 ACM/IEEE
  International Conference for High Performance Computing, Networking, Storage,
  and Analysis}} (\bibinfo{publisher}{IEEE Computer Society},
  \bibinfo{address}{Washington, DC}, \bibinfo{year}{2010}),
  p.~\bibinfo{pages}{1}.

\bibitem[{\citenamefont{Kune\ifmmode~\check{s}\else \v{s}\fi{}
  et~al.}(2007)\citenamefont{Kune\ifmmode~\check{s}\else \v{s}\fi{}, Anisimov,
  Lukoyanov, and Vollhardt}}]{kunes07}
\bibinfo{author}{\bibfnamefont{J.}~\bibnamefont{Kune\ifmmode~\check{s}\else
  \v{s}\fi{}}}, \bibinfo{author}{\bibfnamefont{V.~I.} \bibnamefont{Anisimov}},
  \bibinfo{author}{\bibfnamefont{A.~V.} \bibnamefont{Lukoyanov}},
  \bibnamefont{and}
  \bibinfo{author}{\bibfnamefont{D.}~\bibnamefont{Vollhardt}},
  \bibinfo{journal}{Phys. Rev. B} \textbf{\bibinfo{volume}{75}},
  \bibinfo{pages}{165115} (\bibinfo{year}{2007}).

\bibitem[{\citenamefont{Seth et~al.}()\citenamefont{Seth, Krivenko, Ferrero,
  and Parcollet}}]{sk2015}
\bibinfo{author}{\bibfnamefont{P.}~\bibnamefont{Seth}},
  \bibinfo{author}{\bibfnamefont{I.}~\bibnamefont{Krivenko}},
  \bibinfo{author}{\bibfnamefont{M.}~\bibnamefont{Ferrero}}, \bibnamefont{and}
  \bibinfo{author}{\bibfnamefont{O.}~\bibnamefont{Parcollet}},
  \bibinfo{note}{arXiv:1507.00175}.

\bibitem[{\citenamefont{Parcollet et~al.}(2015)\citenamefont{Parcollet,
  Ferrero, Ayral, Hafermann, Krivenko, Messio, and Seth}}]{pf2015}
\bibinfo{author}{\bibfnamefont{O.}~\bibnamefont{Parcollet}},
  \bibinfo{author}{\bibfnamefont{M.}~\bibnamefont{Ferrero}},
  \bibinfo{author}{\bibfnamefont{T.}~\bibnamefont{Ayral}},
  \bibinfo{author}{\bibfnamefont{H.}~\bibnamefont{Hafermann}},
  \bibinfo{author}{\bibfnamefont{I.}~\bibnamefont{Krivenko}},
  \bibinfo{author}{\bibfnamefont{L.}~\bibnamefont{Messio}}, \bibnamefont{and}
  \bibinfo{author}{\bibfnamefont{P.}~\bibnamefont{Seth}},
  \bibinfo{journal}{Comput. Phys. Commun.} \textbf{\bibinfo{volume}{196}},
  \bibinfo{pages}{398} (\bibinfo{year}{2015}).

\bibitem[{\citenamefont{Sato et~al.}(2015)\citenamefont{Sato, Shirakawa, and
  Yunoki}}]{ss2015}
\bibinfo{author}{\bibfnamefont{T.}~\bibnamefont{Sato}},
  \bibinfo{author}{\bibfnamefont{T.}~\bibnamefont{Shirakawa}},
  \bibnamefont{and} \bibinfo{author}{\bibfnamefont{S.}~\bibnamefont{Yunoki}},
  \bibinfo{journal}{Phys. Rev. B} \textbf{\bibinfo{volume}{91}},
  \bibinfo{pages}{125122} (\bibinfo{year}{2015}).

\bibitem[{\citenamefont{Jarrell and Gubernatis}(1996)}]{MEM}
\bibinfo{author}{\bibfnamefont{M.}~\bibnamefont{Jarrell}} \bibnamefont{and}
  \bibinfo{author}{\bibfnamefont{J.}~\bibnamefont{Gubernatis}},
  \bibinfo{journal}{Physics Reports} \textbf{\bibinfo{volume}{269}},
  \bibinfo{pages}{133 } (\bibinfo{year}{1996}).

\bibitem[{\citenamefont{Schrieffer and Wolff}(1966)}]{sw}
\bibinfo{author}{\bibfnamefont{J.~R.} \bibnamefont{Schrieffer}}
  \bibnamefont{and} \bibinfo{author}{\bibfnamefont{P.~A.} \bibnamefont{Wolff}},
  \bibinfo{journal}{Phys. Rev.} \textbf{\bibinfo{volume}{149}},
  \bibinfo{pages}{491} (\bibinfo{year}{1966}).

\bibitem[{\citenamefont{Nov\'ak
  et~al.}(2013{\natexlab{b}})\citenamefont{Nov\'ak, Kn\'{i}\v{z}ek,
  Mary\v{s}ko, Jir\'ak, and Kune\v{s}}}]{novak13b}
\bibinfo{author}{\bibfnamefont{P.}~\bibnamefont{Nov\'ak}},
  \bibinfo{author}{\bibfnamefont{K.}~\bibnamefont{Kn\'{i}\v{z}ek}},
  \bibinfo{author}{\bibfnamefont{M.}~\bibnamefont{Mary\v{s}ko}},
  \bibinfo{author}{\bibfnamefont{Z.}~\bibnamefont{Jir\'ak}}, \bibnamefont{and}
  \bibinfo{author}{\bibfnamefont{J.}~\bibnamefont{Kune\v{s}}},
  \bibinfo{journal}{J. Phys.: Condens. Matter} \textbf{\bibinfo{volume}{25}},
  \bibinfo{pages}{446001} (\bibinfo{year}{2013}{\natexlab{b}}).

\bibitem[{\citenamefont{Kune\ifmmode~\check{s}\else \v{s}\fi{} and
  Augustinsk\'y}(2014)}]{kunes14a}
\bibinfo{author}{\bibfnamefont{J.}~\bibnamefont{Kune\ifmmode~\check{s}\else
  \v{s}\fi{}}} \bibnamefont{and}
  \bibinfo{author}{\bibfnamefont{P.}~\bibnamefont{Augustinsk\'y}},
  \bibinfo{journal}{Phys. Rev. B} \textbf{\bibinfo{volume}{90}},
  \bibinfo{pages}{235112} (\bibinfo{year}{2014}).

\bibitem[{\citenamefont{Kune\ifmmode~\check{s}\else
  \v{s}\fi{}}(2014)}]{kunes14b}
\bibinfo{author}{\bibfnamefont{J.}~\bibnamefont{Kune\ifmmode~\check{s}\else
  \v{s}\fi{}}}, \bibinfo{journal}{Phys. Rev. B} \textbf{\bibinfo{volume}{90}},
  \bibinfo{pages}{235140} (\bibinfo{year}{2014}).

\bibitem[{\citenamefont{Park et~al.}(2011)\citenamefont{Park, Haule, and
  Kotliar}}]{park11}
\bibinfo{author}{\bibfnamefont{H.}~\bibnamefont{Park}},
  \bibinfo{author}{\bibfnamefont{K.}~\bibnamefont{Haule}}, \bibnamefont{and}
  \bibinfo{author}{\bibfnamefont{G.}~\bibnamefont{Kotliar}},
  \bibinfo{journal}{Phys. Rev. Lett.} \textbf{\bibinfo{volume}{107}},
  \bibinfo{pages}{137007} (\bibinfo{year}{2011}).

\bibitem[{\citenamefont{Boehnke and Lechermann}(2012)}]{boehnke12}
\bibinfo{author}{\bibfnamefont{L.}~\bibnamefont{Boehnke}} \bibnamefont{and}
  \bibinfo{author}{\bibfnamefont{F.}~\bibnamefont{Lechermann}},
  \bibinfo{journal}{Phys. Rev. B} \textbf{\bibinfo{volume}{85}},
  \bibinfo{pages}{115128} (\bibinfo{year}{2012}).

\end{thebibliography}
\end{document}